\documentclass[amsmath,amssymb,superscriptaddress,nofootinbib,11pt,a4paper]{revtex4}

\usepackage[utf8]{inputenc}
\usepackage[T1]{fontenc}
\usepackage{xcolor}
\usepackage{amsfonts}
\usepackage{graphicx} 
\usepackage{dcolumn}   
\usepackage{bm}
\usepackage{float}
\usepackage{braket}
\usepackage{mathrsfs}
\usepackage{subcaption} 
\usepackage{siunitx}
\usepackage{comment}
\usepackage{diagbox} 
\usepackage{textcomp}
\usepackage{epstopdf}
\usepackage{wrapfig}

\begin{document}
\title{Detecting Frame Dragging Effect with Next-Generation Spaceborne Clocks in a LEO–DRO Satellite Configuration}

\author{Li-Fang Li}
\email{lilifang@imech.ac.cn}
\affiliation{Center for Gravitational Wave Experiment, National Microgravity Laboratory, Institute of Mechanics, Chinese Academy of Sciences, Beijing 100190, China}

\author{Hanlin Song}
\email{songhanlin@imech.ac.cn}
\affiliation{Center for Gravitational Wave Experiment, National Microgravity Laboratory, Institute of Mechanics, Chinese Academy of Sciences, Beijing 100190, China}

\author{Xing Bian}
\affiliation{Center for Gravitational Wave Experiment, National Microgravity Laboratory, Institute of Mechanics, Chinese Academy of Sciences, Beijing 100190, China}

\author{Tong Liu}
\email{liutong2021@csu.ac.cn}
\affiliation{Technology and Engineering Center for Space Utilization, Chinese Academy of Science}

\author{Shao-Jiang Wang}
\email{schwang@itp.ac.cn}
\affiliation{Institute of Theoretical Physics, Chinese Academy of Sciences (CAS), Beijing 100190, China}
\affiliation{Asia Pacific Center for Theoretical Physics (APCTP), Pohang 37673, Korea
}

\author{Song He}
\email{hesong@nbu.edu.cn}
\affiliation{Institute of Fundamental Physics and Quantum Technology,
 Department of Physics, School of Physical Science and Technology,
 Ningbo University, Ningbo, Zhejiang 315211, China}

\author{Yang Li}
\affiliation{Technology and Engineering Center for Space Utilization, Chinese Academy of Science}
\affiliation{University of Chinese Academy of Sciences}

\author{Jiaxing Leng}
\affiliation{Innovation Academy for Microsatellites, Chinese Academy of Sciences, No.1 Xueyang Road, Pudong New Area, Shanghai, 201304}

\author{Ziren Luo}
\affiliation{Center for Gravitational Wave Experiment, National Microgravity Laboratory, Institute of Mechanics, Chinese Academy of Sciences, Beijing 100190, China}
\affiliation{Key Laboratory of Gravitational Wave Precision Measurement of Zhejiang Province, Hangzhou Institute for Advanced Study, UCAS, Hangzhou 310024, China}

\author{Peng Xu}
\email{xupeng@imech.ac.cn}
\affiliation{Center for Gravitational Wave Experiment, National Microgravity Laboratory, Institute of Mechanics, Chinese Academy of Sciences, Beijing 100190, China}
\affiliation{Taiji Laboratory for Gravitational Wave Universe (Beijing/Hangzhou), University of Chinese Academy of Sciences, Beijing 100049, China}
\affiliation{Hangzhou Institute for Advanced Study, University of Chinese Academy of Sciences, Hangzhou 310024, China}

\author{Rong-Gen Cai}
 \email{caironggen@nbu.edu.cn}
\affiliation{Institute of Fundamental Physics and Quantum Technology,
 Department of Physics, School of Physical Science and Technology,
 Ningbo University, Ningbo, Zhejiang 315211, China}

\begin{abstract}
Our approach employs a satellite constellation comprising one satellite in low Earth orbit (LEO) and one satellite in distant retrograde orbit (DRO).
This configuration is designed to enhance the sensitivity to the Earth's rotation-induced frame-dragging signal while providing a stable and gravitationally isolated frequency reference. We derive a complete relativistic formulation for the frequency comparison between LEO and DRO, incorporating the post-Newtonian metric, higher-order Earth multipoles, lunar and solar tidal potentials, and the gravitomagnetic vector potential. The frame-dragging signal is of order $10^{-20}$, a level that is becoming increasingly accessible with next-generation optical clocks. Recent advances in zero-dead-time strontium optical lattice clocks have demonstrated a fractional frequency instability of $2.9\times10^{-19}$ at an averaging time of 20,000~s, with an estimated one-day stability of $2.5\times10^{-19}$. 
With continued advances in clock stability, achieving fractional-frequency sensitivity at the $10^{-20}$ level appears feasible in the near term, potentially enabling the direct measurement of a frame-dragging signal of the same order.
Such a measurement would constrain Earth's gravitomagnetic field, enabling future investigations of gravitomagnetic effects on Einstein clock synchronization.
\end{abstract}

\maketitle

\section{Introduction}
\label{sec:introduction}

Optical atomic clocks have reached fractional frequency stabilities
beyond $1\times10^{-16}/\sqrt{\tau}$ and systematic uncertainties
approaching $10^{-19}$~\cite{Bloom2014, McGrew2018}, and are now being
transitioned from laboratory instruments into space-qualified payloads.
Recent advances in optical precision timing include the Robust Optical Clock Network program of the U.S. Defense Advanced Research Projects Agency \cite{Roslund:2024nak, Hilton2025}, the European Space Agency's  Optical Synchronised Time and Ranging in-orbit demonstrator  \cite{Blonski2025}, the German Aerospace Center's COMPASSO mission for space-based optical clocks and laser-link technologies \cite{Kuschewski2024}, and the successful in-orbit demonstration of an optical frequency comb by Australia's QuantX Labs \cite{Ahern2025}. Together with an international comparison involving ten optical clocks across six countries \cite{Lindvall2025}, these developments underscore significant progress toward space-based optical timekeeping and globally interconnected optical-clock networks.

Recent developments in optical clocks and precision time transfer in China have paved the way for their potential applications in space.
The National Time Service Center 
has demonstrated picosecond-level time transfer over a 2061-km optical-fiber link,  achieving a time deviation of 5.6~ps at an averaging time of 1~s~\cite{Guo2025}.
Its strontium optical clock, NTSC-Sr2, has achieved both fractional frequency instability and systematic uncertainty below $2\times10^{-18}$~\cite{NTSCSr2_2025, Xu2026}. 
The Beijing
Satellite Navigation Center has demonstrated 160~days of autonomous
time-keeping with a calcium-ion optical clock~\cite{Li2025}. In space,
the DRO-A/B/L three-satellite constellation, the world's first
DRO-based cislunar constellation, has demonstrated inter-satellite
time-frequency transfer and ranging over distances reaching $1.17\times10^6$~km~\cite{Leng2025,Luo2026,Zhang2025,Chen2022, Liu2023}, 
and the Mengtian module of the China Space Station carries a cold-atom optical clock~\cite{Liu2023, Zhang2024}. 
These advances provide a technological basis for future space-based precision measurements and tests of fundamental physics.

Among the most compelling objectives for such clocks is the detection of
the frame-dragging effect. Its precise measurement constitutes a
cornerstone test of general relativity in the weak-field, slow-motion
regime~\cite{Ciufolini2004}.
Satellite laser ranging observations of LAGEOS and LAGEOS~II achieved approximately 10\% precision~\cite{Ciufolini2004}. The Gravity Probe B mission provided a complementary measurement but was limited to approximately 20\% precision by unmodeled systematic effects~\cite{Everitt2011}. The ongoing LAGEOS--LARES~2 experiment aims to improve the measurement precision by at least one order of magnitude~\cite{Ciufolini2023, Iorio2025}.
Despite these pioneering efforts, achieving further improvements in precision while maintaining robust control of systematic uncertainties remains a significant experimental challenge.

The key difference between our proposed approach and previous satellite-based tests of frame dragging lies in the underlying measurement principle.
LAGEOS/LAGEOS~II/LARES~2 infer the gravitomagnetic field from orbital precession
\cite{Ciufolini2004, Ciufolini2023, Iorio2025}, while Gravity Probe B from
the precession of a cryogenic gyroscope~\cite{Everitt2011}. 
Our scheme instead exploits the frequency comparison of two
optical clocks, one in a high-inclination LEO orbit and the other in DRO.
The frequency comparison measures their relative proper time rates,
from which the gravitomagnetic frequency shift arising from the $g_{0i}$
components of the metric can be extracted after subtracting
gravitational and kinematic backgrounds. This provides a direct,
clock-based test of general relativity, conceptually related to the
gravitomagnetic clock effect~\cite{Tartaglia2000, Wassermann2025}, with systematic uncertainties distinct from those of conventional precession-based experiments. 
Recent progress towards
$10^{-19}$-level clocks~\cite{Bloom2014, McGrew2018}, including a
multiplexed optical lattice clock with a fractional instability of
$9.7(4)\times10^{-18}/\sqrt{\tau}$ and an uncertainty of
$8.9\times10^{-20}$ after 3.3~hours~\cite{Zheng2022}, indicates that the
clock quantum projection noise can fall below the frame-dragging signal
in the coming decade.

Building on these advances, we propose a two-satellite mission concept to probe terrestrial frame dragging through high-precision optical-clock comparisons between a high-inclination low-Earth-orbit (LEO) satellite and a satellite in a lunar distant retrograde orbit (DRO). The two clocks are  
connected by a bidirectional optical frequency-transfer link, and the gravitomagnetic frequency shift is
extracted after subtracting the modeled gravitational backgrounds. 
The high-inclination LEO orbit provides a measurable frame-dragging contribution, while the DRO provides a dynamically stable frequency reference in a well-characterized gravitational environment.
An appropriately constructed two-way frequency observable suppresses the leading first-order Doppler contribution, and the differential
measurement suppresses common-mode gravitational and instrumental
systematics. Unlike previous proposals that relied on geostationary or
high-Earth-orbit references, this LEO--DRO architecture draws directly
on the operational experience of the DRO constellation and bridges
quantum metrology, satellite infrastructure, and fundamental physics.

The paper is organized as follows. Section II presents the mission architecture for LEO-DRO optical clock comparison, describing the satellite constellation, the bidirectional phase-stabilised laser link, and the key hardware specifications. Section III establishes the theoretical framework and the frequency ratio decomposition, and then provides a detailed error analysis and suppression strategies. Finally, Section IV concludes the paper with a summary of the proposed mission and an outlook for future improvements in optical clock stability and relativistic gravimetry.

\section{MISSION ARCHITECTURE OVERVIEW FOR LEO-DRO OPTICAL CLOCK COMPARISON}

\begin{figure}[htbp]
\centering
\includegraphics[width=0.85\textwidth]{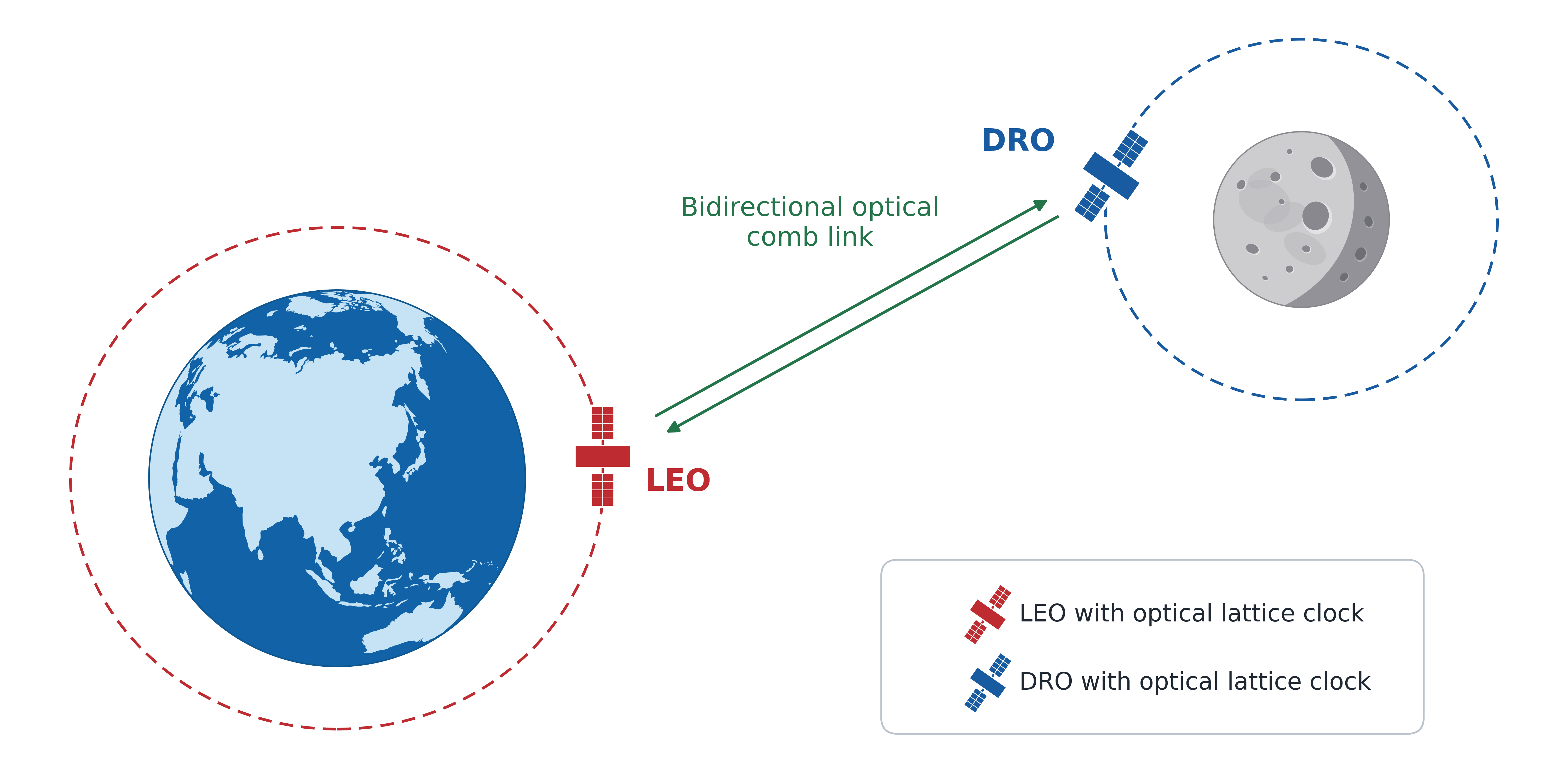}
\caption{Schematic of the proposed LEO-DRO constellation for frame-dragging detection. The system comprises a high-inclination low Earth orbit (LEO) satellite and one distant retrograde orbit (DRO) satellite at lunar distance, all equipped with optical lattice clocks and interconnected by bidirectional pulsed laser links based on dual optical frequency combs.The LEO satellite has an orbital angular momentum nearly orthogonal to Earth's rotation axis, while the DRO satellite provides a frequency reference in a region of relatively weak gravitational potential.  The bidirectional laser links enable common-mode Doppler cancellation and differential gravitational potential comparison, isolating the tiny gravitomagnetic frequency shift from the dominant Earth-centered redshift. The inset illustrates the $2:1$ resonance between the DRO and the Moon's orbital period.}
\label{constellation}
\end{figure}

Building on previous developments in spaceborne optical clocks and DRO satellite missions, this section presents the mission architecture designed for detecting the frame-dragging effect. 
We first describe the constellation consisting of one LEO satellite and one DRO satellite, then introduce the bidirectional laser link for suppressing the first-order Doppler shift, and finally summarize the key hardware requirements.
The overall constellation concept is illustrated in Fig.~\ref{constellation}.

The proposed constellation consists of one LEO satellite and one DRO satellite. The LEO satellite is placed in a near-circular, high-inclination orbit at an altitude of $\sim 630$ km, corresponding to a geocentric radius of $\sim 7000$ km. This choice is motivated by two considerations. First, the frame-dragging frequency shift scales as $r^{-\frac{5}{2}}$ and is maximized at low altitudes. 
Second, an inclined orbit is adopted to retain a nonzero projection of the orbital angular momentum onto Earth's spin axis, thereby preserving a measurable frame-dragging signal while still mitigating the contamination from Earth's complex higher-order gravitational harmonics. 
Small deviations from the nominal orbital inclination modify the frame-dragging contribution and must be accounted for in the orbit and frequency-transfer models. At the nominal LEO altitude, the orbital velocity is approximately $7.55$ km/s, yielding an orbital period of about $97$ minutes, which sets the characteristic timescale for the satellite orbital motion.

The DRO satellite is placed in a distant retrograde orbit with a geocentric distance ranging from approximately $300{,}000$ km to $450{,}000$ km, at the boundary of the Earth's gravitational sphere of influence. 
The DRO belongs to a family of periodic orbits in the Earth--Moon three-body system. In the Earth--Moon rotating frame, the spacecraft moves retrogradely around the Moon, whereas its overall geocentric motion is prograde in the Earth-centered inertial frame. We consider a 2:1 resonant DRO, for which the spacecraft completes approximately two orbital cycles in the rotating frame during one lunar orbital period. Such orbits can exhibit long-term dynamical stability, although their actual evolution is affected by solar perturbations, lunar orbital eccentricity, and other non-Keplerian effects.
The gravitational fields of the Earth and Moon are balanced in this region, creating a stable gravitational environment with minimal tidal variations. The orbital velocity of the DRO satellite is approximately $1.3$ km/s, significantly lower than that of LEO, which reduces the kinematic Doppler shift to be cancelled by the bidirectional link. The DRO satellite serves as the frequency reference, providing a well-characterized and gravitationally isolated clock signal against which the LEO clock is compared.

The frequency comparison between LEO and DRO is performed via a bidirectional pulsed laser link based on dual optical frequency combs, operating in the near-infrared. 
The proposed system follows a measurement concept analogous to the dual one-way ranging (DOWR) technique demonstrated by the DRO-A mission, but employs optical frequency-comb signals instead of K-band microwave signals, providing two to three orders of magnitude higher stability and precision. 
The principle of the bidirectional link is as follows. Each satellite transmits a phase-stabilized optical pulse train toward the other satellite.  At each receiving terminal, the arrival time of the incoming optical pulse train is detected and compared with the local time reference provided by the onboard optical frequency comb. By transmitting in both directions simultaneously and subtracting the two measurements, the opposite-signed Doppler contributions from the relative motion of the two satellites cancel at the first order. In practice, residual errors arise from asymmetric delays in the uplink and downlink paths, including fiber dispersion, optical component temperature drifts, and relativistic Shapiro time delay variations. These residual effects are suppressed by active timing stabilisation, which uses a low-bandwidth auxiliary link to measure the instantaneous Doppler shift and pre-distort the outgoing optical phase via a fiber-optic delay line or an acousto-optic modulator, achieving a noise suppression ratio exceeding $40$ dB below $1$ Hz. 

Each satellite's laser terminal comprises a telescope with a $15$ cm aperture, a fast steering mirror for fine pointing control with residual jitter below $0.1$ $\mu$rad, and a pointing-ahead mechanism to compensate for the light travel time between LEO and DRO (approximately $1.0$--$1.5$ s). The received optical pulses are detected by low-noise InGaAs photoreceivers, and their arrival times are recorded using high-precision event timers with a resolution better than $0.5$ ps. The two-way time-of-flight comparison yields the fractionalfrequency ratio $f_{\mathrm{LEO}}/f_{\mathrm{DRO}}$ in real time. The entire link is referenced to the onboard optical frequency comb, ensuring that the combined measurement uncertainty from the time-frequency transfer remains below $1\times10^{-18}$ after one day of averaging.

Each satellite carries an identical optical lattice clock based on either $^{87}$Sr or $^{171}$Yb atoms. The clock achieves a fractional frequency instability $\sigma_y(\tau) \leqslant 1\times10^{-16}/\sqrt{\tau}$, with a projected daily drift below $1\times10^{-18}$. The clock subsystem includes a compact atomic source maintaining ultra-high vacuum below $10^{-8}$ Pa, a magic-wavelength lattice laser (813 nm for Sr or 759 nm for Yb) with sub-hertz linewidth, and an ultra-stable clock laser (698 nm for Sr or 578 nm for Yb) pre-stabilized to a high-finesse cavity made of single-crystal silicon or ultra-low-expansion glass, with active thermal control maintaining the cavity temperature stable to $\pm 0.5$ mK. A fiber-based femtosecond optical frequency comb generates the pulsed optical carrier for inter-satellite time transfer, and also phase-coherently bridges the optical clock frequency to the microwave domain for onboard reference distribution.

The satellite platforms are designed to minimize environmental disturbances affecting the onboard optical clocks.
A drag-free control system based on high-sensitivity accelerometers (noise floor $< 3\times10^{-12}$ m/s$^2/\sqrt{\mathrm{Hz}}$) and cold-gas or field-emission electric thrusters maintains residual non-gravitational acceleration below $1\times10^{-9}$ m/s$^2$ in the $1$ mHz to $0.1$ Hz band. Multi-layer $\mu$-metal shielding reduces the ambient magnetic field to less than $1$ nT, suppressing the quadratic Zeeman shift. For precise orbit determination, the LEO satellite carries a triple-frequency GNSS receiver (GPS/Galileo/BeiDou) for real-time kinematic positioning, while all satellites are equipped with corner-cube retroreflectors for ground-based satellite laser ranging, achieving centimetre to sub-centimetre orbital accuracy. 
Since GNSS coverage is limited at lunar distances, the DRO satellite relies on inter-satellite ranging, together with dynamical orbit propagation and supplementary tracking information, for orbit determination.

With further improvements in the proposed hardware, the LEO--DRO configuration could provide the sensitivity required to resolve a frame-dragging frequency shift of order $10^{-20}$, provided that the residual systematic uncertainties are sufficiently controlled. We now proceed to establish the theoretical framework that relates the measured frequency ratio to the underlying relativistic effects, including the gravitomagnetic frame-dragging term.

\begin{figure}[htbp]
\centering
\includegraphics[width=0.7\textwidth]{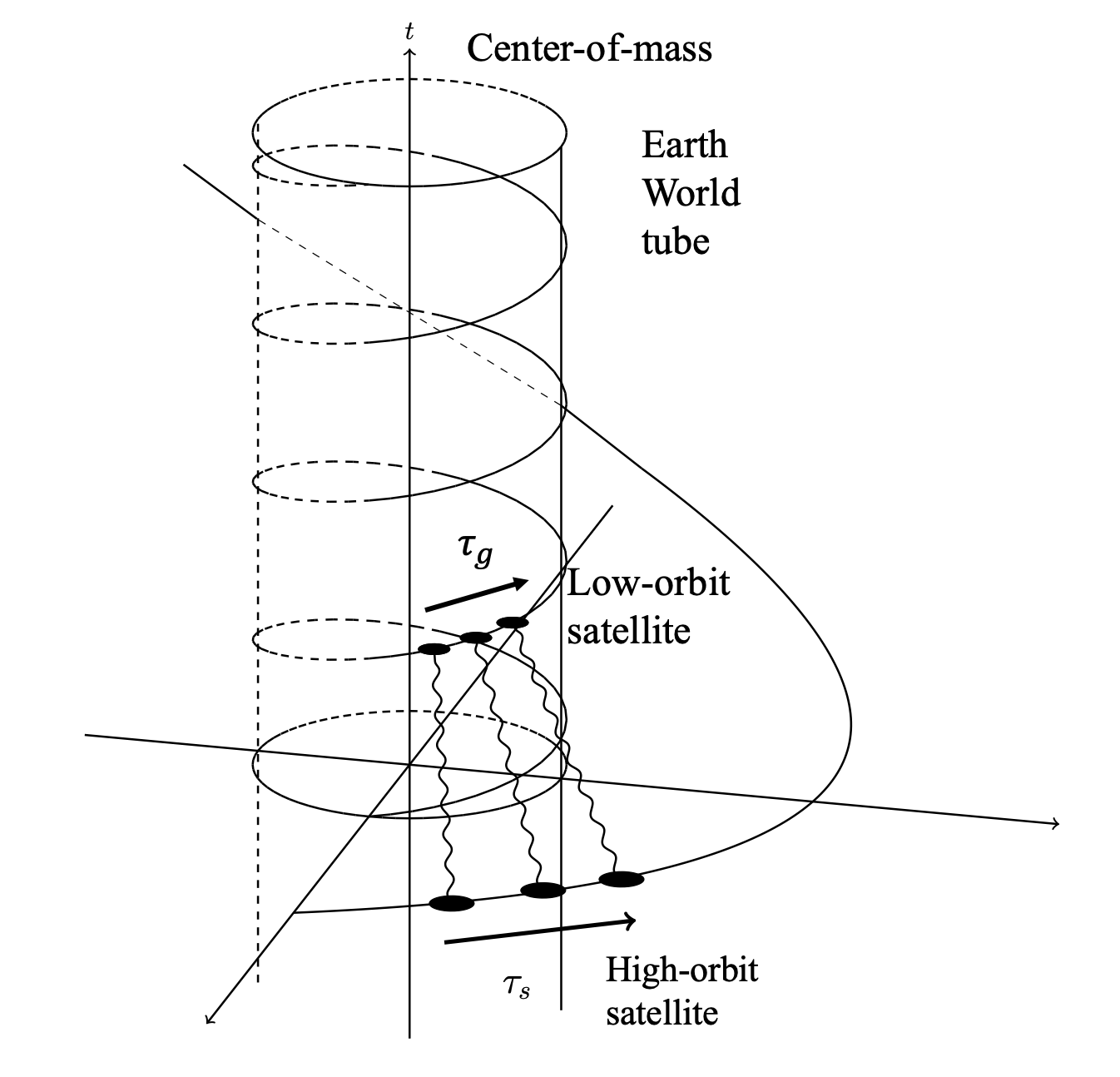}
\caption{Relativistic gravimetry via time-frequency comparison.}
\label{fig:freqcomparison}
\end{figure}

With the mission architecture established in the preceding section, we now develop a rigorous relativistic model for the frequency comparison between the LEO and DRO satellites. The concept of the proposed LEO-DRO frequency comparison is illustrated in Fig.~\ref{fig:freqcomparison}. The observed fractional frequency difference is governed by the gravitational redshift and kinematic effects predicted by general relativity. To compute the expected signal and to isolate the tiny frame-dragging contribution, we derive the frequency ratio from first principles using the Geocentric Celestial Reference System (GCRS). The GCRS provides a consistent post-Newtonian framework for describing satellite motion and time transfer in the vicinity of Earth, incorporating Earth's gravitational multipoles, lunar and solar tidal potentials, and gravitomagnetic effects due to rotating masses.

In the GCRS with coordinates $X^{\mu} = (cT, \mathbf{X})$, the spacetime metric in the post-Newtonian framework is given by~\cite{Soffel2003,Brumberg2008}:
\begin{align}
g_{00} &= -1 + \frac{2W}{c^2} - \frac{2W^2}{c^4} + \mathcal{O}(c^{-5}), \label{eq:metric00} \\
g_{0i} &= -\frac{4}{c^3} W_i + \mathcal{O}(c^{-5}), \label{eq:metric0i} \\
g_{ij} &= \delta_{ij}\left(1 + \frac{2W}{c^2}\right) + \mathcal{O}(c^{-4}), \label{eq:metricij}
\end{align}
where $W(T,\mathbf{X})$ is the total scalar gravitational potential (Earth + tides) and $W_i(T,\mathbf{X})$ is the vector potential. The proper time $\tau$ of a clock moving with coordinate velocity $\mathbf{v} = d\mathbf{X}/dT$ is obtained from
\begin{equation}
d\tau = \frac{1}{c}\sqrt{-g_{\mu\nu}\frac{dX^{\mu}}{dT}\frac{dX^{\nu}}{dT}}\,dT. \label{eq:propertime}
\end{equation}
Here $\tau$ denotes the proper time measured by the onboard optical clock along the satellite's worldline, while $T$ is the Geocentric Coordinate Time (TCG) in the GCRS, as shown in Fig.~\ref{fig:freqcomparison}.
For a satellite in a weak gravitational field and with slow motion $(v \ll c)$, the metric can be expanded. To maintain formal completeness and to ensure that all potentially relevant effects are included, we write the LEO and DRO metrics in their full forms, keeping terms up to $\mathcal{O}(c^{-4})$. In spherical coordinates $(r, \theta, \phi)$, the LEO metric reads~\cite{Soffel2003,Brumberg2008}:
\begin{align}
g_{00}^{\mathrm{LEO}} = & -1 + \frac{2}{c^2} \Bigg[ \frac{GM_{\oplus}}{r} 
+ \frac{GM_{\mathrm{Moon}}}{|\mathbf{r} - \mathbf{r}_{\mathrm{Moon}}|} 
+ \frac{GM_{\odot}}{|\mathbf{r} - \mathbf{r}_{\odot}|} \nonumber \\
& \quad + \frac{GM_{\oplus}}{r} \sum_{l=2}^{\infty} \left( \frac{R_{\oplus}}{r} \right)^l \sum_{m=-l}^{l} C_{lm} Y_{lm}(\theta, \phi) \Bigg] \nonumber \\
& - \frac{2}{c^4} \Bigg[ \frac{G^2 M_{\oplus}^2}{r^2} 
+ \frac{G^2 M_{\oplus} M_{\mathrm{Moon}}}{|\mathbf{r} - \mathbf{r}_{\mathrm{Moon}}| r} 
+ \frac{G^2 M_{\oplus} M_{\odot}}{|\mathbf{r} - \mathbf{r}_{\odot}| r} \Bigg] + \mathcal{O}(c^{-5}), \label{eq:LEOmetric00} \\[6pt]
g_{0\phi}^{\mathrm{LEO}} = & -\frac{2G}{c^3} \frac{J_{\oplus} \sin^2\theta}{r}
- \frac{2G}{c^3} \frac{J_{\mathrm{Moon}} \sin^2\theta}{|\mathbf{r} - \mathbf{r}_{\mathrm{Moon}}|}
- \frac{2G}{c^3} \frac{J_{\odot} \sin^2\theta}{|\mathbf{r} - \mathbf{r}_{\odot}|}
+ \mathcal{O}(c^{-5}), \label{eq:LEOmetric0phi} \\[6pt]
g_{rr}^{\mathrm{LEO}} = & 1 + \frac{2}{c^2} \frac{GM_{\oplus}}{r}
+ \frac{2}{c^2} \frac{GM_{\oplus}}{r} \sum_{l=2}^{\infty} \left( \frac{R_{\oplus}}{r} \right)^l \sum_{m=-l}^{l} C_{lm} Y_{lm}(\theta, \phi) \nonumber \\
& + \frac{2}{c^2} \frac{GM_{\mathrm{Moon}}}{|\mathbf{r} - \mathbf{r}_{\mathrm{Moon}}|}
+ \frac{2}{c^2} \frac{GM_{\odot}}{|\mathbf{r} - \mathbf{r}_{\odot}|} + \mathcal{O}(c^{-4}), \label{eq:LEOmetricrr} \\[4pt]
g_{\theta\theta}^{\mathrm{LEO}} = & r^2 \Bigg[ 1 + \frac{2}{c^2} \frac{GM_{\oplus}}{r}
+ \frac{2}{c^2} \frac{GM_{\oplus}}{r} \sum_{l=2}^{\infty} \left( \frac{R_{\oplus}}{r} \right)^l \sum_{m=-l}^{l} C_{lm} Y_{lm}(\theta, \phi) \nonumber \\
& + \frac{2}{c^2} \frac{GM_{\mathrm{Moon}}}{|\mathbf{r} - \mathbf{r}_{\mathrm{Moon}}|}
+ \frac{2}{c^2} \frac{GM_{\odot}}{|\mathbf{r} - \mathbf{r}_{\odot}|} \Bigg] + \mathcal{O}(c^{-4}), \label{eq:LEOmetrictwotheta} \\[4pt]
g_{\phi\phi}^{\mathrm{LEO}} = & r^2 \sin^2\theta \Bigg[ 1 + \frac{2}{c^2} \frac{GM_{\oplus}}{r}
+ \frac{2}{c^2} \frac{GM_{\oplus}}{r} \sum_{l=2}^{\infty} \left( \frac{R_{\oplus}}{r} \right)^l \sum_{m=-l}^{l} C_{lm} Y_{lm}(\theta, \phi) \nonumber \\
& + \frac{2}{c^2} \frac{GM_{\mathrm{Moon}}}{|\mathbf{r} - \mathbf{r}_{\mathrm{Moon}}|}
+ \frac{2}{c^2} \frac{GM_{\odot}}{|\mathbf{r} - \mathbf{r}_{\odot}|} \Bigg] + \mathcal{O}(c^{-4}), \label{eq:LEOmetricphiphi}
\end{align}
where $C_{lm}$ are the fully normalized spherical harmonic coefficients, $Y_{lm}$ the spherical harmonics, and $J_{\oplus}$, $J_{\mathrm{Moon}}$, $J_{\odot}$ are the angular momentum parameters for the Earth, Moon's orbital motion, and Sun's orbital motion, respectively. The DRO metric takes a similar form, with the Earth's spherical harmonic expansion truncated at $l=4$ due to the rapid decay of higher-order multipoles at lunar distances:
\begin{align}
g_{00}^{\mathrm{DRO}} = & -1 + \frac{2}{c^2} \Bigg[ \frac{GM_{\oplus}}{r} 
+ \frac{GM_{\mathrm{Moon}}}{|\mathbf{r} - \mathbf{r}_{\mathrm{Moon}}|} 
+ \frac{GM_{\odot}}{|\mathbf{r} - \mathbf{r}_{\odot}|} \nonumber \\
& \quad + \frac{GM_{\oplus}}{r} \sum_{l=2}^{4} \left( \frac{R_{\oplus}}{r} \right)^l \sum_{m=-l}^{l} C_{lm} Y_{lm}(\theta, \phi) \Bigg] \nonumber \\
& - \frac{2}{c^4} \Bigg[ \frac{G^2 M_{\oplus}^2}{r^2} 
+ \frac{G^2 M_{\oplus} M_{\mathrm{Moon}}}{|\mathbf{r} - \mathbf{r}_{\mathrm{Moon}}| r} 
+ \frac{G^2 M_{\oplus} M_{\odot}}{|\mathbf{r} - \mathbf{r}_{\odot}| r} \Bigg] + \mathcal{O}(c^{-5}), \label{eq:DROmetric00} \\[6pt]
g_{0\phi}^{\mathrm{DRO}} = & -\frac{2G}{c^3} \frac{J_{\oplus} \sin^2\theta}{r}
- \frac{2G}{c^3} \frac{J_{\mathrm{Moon}} \sin^2\theta}{|\mathbf{r} - \mathbf{r}_{\mathrm{Moon}}|}
- \frac{2G}{c^3} \frac{J_{\odot} \sin^2\theta}{|\mathbf{r} - \mathbf{r}_{\odot}|}
+ \mathcal{O}(c^{-5}), \label{eq:DROmetric0phi} \\[6pt]
g_{rr}^{\mathrm{DRO}} = & 1 + \frac{2}{c^2} \frac{GM_{\oplus}}{r}
+ \frac{2}{c^2} \frac{GM_{\oplus}}{r} \sum_{l=2}^{4} \left( \frac{R_{\oplus}}{r} \right)^l \sum_{m=-l}^{l} C_{lm} Y_{lm}(\theta, \phi) \nonumber \\
& + \frac{2}{c^2} \frac{GM_{\mathrm{Moon}}}{|\mathbf{r} - \mathbf{r}_{\mathrm{Moon}}|}
+ \frac{2}{c^2} \frac{GM_{\odot}}{|\mathbf{r} - \mathbf{r}_{\odot}|} + \mathcal{O}(c^{-4}), \label{eq:DROmetricrr} \\[4pt]
g_{\theta\theta}^{\mathrm{DRO}} = & r^2 \Bigg[ 1 + \frac{2}{c^2} \frac{GM_{\oplus}}{r}
+ \frac{2}{c^2} \frac{GM_{\oplus}}{r} \sum_{l=2}^{4} \left( \frac{R_{\oplus}}{r} \right)^l \sum_{m=-l}^{l} C_{lm} Y_{lm}(\theta, \phi) \nonumber \\
& + \frac{2}{c^2} \frac{GM_{\mathrm{Moon}}}{|\mathbf{r} - \mathbf{r}_{\mathrm{Moon}}|}
+ \frac{2}{c^2} \frac{GM_{\odot}}{|\mathbf{r} - \mathbf{r}_{\odot}|} \Bigg] + \mathcal{O}(c^{-4}), \label{eq:DROmetrictwotheta} \\[4pt]
g_{\phi\phi}^{\mathrm{DRO}} = & r^2 \sin^2\theta \Bigg[ 1 + \frac{2}{c^2} \frac{GM_{\oplus}}{r}
+ \frac{2}{c^2} \frac{GM_{\oplus}}{r} \sum_{l=2}^{4} \left( \frac{R_{\oplus}}{r} \right)^l \sum_{m=-l}^{l} C_{lm} Y_{lm}(\theta, \phi) \nonumber \\
& + \frac{2}{c^2} \frac{GM_{\mathrm{Moon}}}{|\mathbf{r} - \mathbf{r}_{\mathrm{Moon}}|}
+ \frac{2}{c^2} \frac{GM_{\odot}}{|\mathbf{r} - \mathbf{r}_{\odot}|} \Bigg] + \mathcal{O}(c^{-4}), \label{eq:DROmetricphiphi}
\end{align}
where $R_{\mathrm{EM}} = 3.844\times10^8$ m is the mean Earth--Moon distance, $\omega_{\mathrm{Moon}} = 2.662\times10^{-6}$ rad/s the Moon's orbital angular velocity, and $M_{\odot}$ the solar mass. The vector potential includes contributions from Earth's rotation and the Moon's orbital motion, which affect the DRO through the total angular momentum $J_{\mathrm{tot}} = J_{\oplus} + J_{\mathrm{Moon}}$.

Since frequencies are proportional to the inverse of periods and periods are proportional to proper times, the frequency ratio \(f_{\mathrm{LEO}}/f_{\mathrm{DRO}}\),  denoting the ratio of the intrinsic proper frequencies of the two optical clocks, can be expressed as
\begin{equation}
\frac{f_{\mathrm{LEO}}}{f_{\mathrm{DRO}}} = \frac{d\tau_{\mathrm{DRO}}}{d\tau_{\mathrm{LEO}}} = \sqrt{\frac{-g_{\mu\nu}^{\mathrm{DRO}} v_{\mathrm{DRO}}^\mu v_{\mathrm{DRO}}^\nu}{-g_{\mu\nu}^{\mathrm{LEO}} v_{\mathrm{LEO}}^\mu v_{\mathrm{LEO}}^\nu}} \times (1 + N),
\label{eq:freqratio}
\end{equation}
where \(v^\mu \equiv (c, \mathbf{v})\) denotes the coordinate four-velocity of the satellite with \(\mathbf{v} = d\mathbf{X}/dT\) being the ordinary three-velocity in the GCRS. Here \(N\) represents the residual relative measurement noise after bidirectional Doppler cancellation and gravitational background subtraction. In practice, the measured frequency ratio obtained from the bidirectional laser link contains propagation and kinematic effects; after subtracting these modeled contributions, the residual approaches the intrinsic ratio defined above.

To isolate gravitational field information from kinematic effects, we perform a Taylor expansion of the square root terms up to second-order perturbations. For a general metric component, this expansion yields:
\begin{align}
\frac{d\tau}{dT} 
&= \sqrt{-g_{00} - 2 \frac{g_{0i}}{c} v^i - \frac{g_{ij}}{c^2} v^i v^j} \nonumber \\
&= 1 - \frac{W}{c^2} - \frac{v^2}{2c^2} 
   + \frac{1}{c^4} \left( \frac{W^2}{2} + {4W v}- \frac{3Wv^2}{2} - \frac{v^4}{8} \right)    + \mathcal{O}(c^{-5})
\label{eq:proper_time_expansion}
\end{align}
For circular orbits, the $v^4/8$ term in the expansion of Eq.~\eqref{eq:proper_time_expansion}  is constant and is absorbed into the static offset. This treatment is appropriate since the target frame-dragging signal itself is a constant offset of order $10^{-20}$, as shown below.

The calculation of the orbital frequency ratio $f_{\mathrm{LEO}}/f_{\mathrm{DRO}}$ for circular orbits is predicated on the circular orbit approximation, which is well-justified for both LEO and DRO missions. Inserting the full LEO and DRO metrics into Eq.~\eqref{eq:freqratio} and including the leading gravitomagnetic contribution, we obtain the complete frequency ratio for circular orbits:
\begin{align}
\frac{f_{\mathrm{LEO}}}{f_{\mathrm{DRO}}} = & 1 + \frac{U_{\mathrm{DRO}} - U_{\mathrm{LEO}}}{c^2} \nonumber \\
& + \frac{2G}{c^4}\left(J_{\mathrm{tot}}\frac{h_z^{\mathrm{DRO}}}{r_{\mathrm{DRO}}^3} - J_{\oplus}\frac{h_z^{\mathrm{LEO}}}{r_{\mathrm{LEO}}^3}\right) \nonumber \\
& + \frac{1}{2c^2}\left[\left(1 + \frac{2GM_{\oplus}}{c^2r_{\mathrm{DRO}}}\right)v_{\mathrm{DRO}}^2 - \left(1 + \frac{2GM_{\oplus}}{c^2r_{\mathrm{LEO}}}\right)v_{\mathrm{LEO}}^2\right] \nonumber \\
& - \frac{2}{c^4}\left(\frac{G^2 M_{\oplus}^2}{r_{\mathrm{DRO}}^2} - \frac{G^2 M_{\oplus}^2}{r_{\mathrm{LEO}}^2}\right) + \mathcal{O}(c^{-5}), \label{eq:freqratio_full}
\end{align}
where $h_z = (\mathbf{r} \times \mathbf{v})_z$ is the $z$-component of the orbital angular momentum per unit mass. It is important to clarify that Eq.~\eqref{eq:freqratio_full} includes only those terms that contribute to the frequency ratio at the level of interest. Constant or slowly varying contributions, such as the Earth--Sun and Earth--Moon nonlinear cross terms at the $\sim10^{-22}$ level, are not explicitly included in this decomposition. These contributions will be modeled and subtracted using DE440 ephemerides and high-resolution geopotential models, as discussed in Sec.~\ref{sec:systematics}.  A complete order-of-magnitude justification for all retained and omitted terms is provided in Appendix~\ref{app:orders}. 

With this truncation, the gravitational potentials \(U_{\mathrm{LEO}}\) and \(U_{\mathrm{DRO}}\) are defined as the sums of all scalar potential contributions (excluding the nonlinear $-2W^2/c^4$ terms which are treated separately in Eq.~\eqref{eq:DeltaNonlinear_def}):
\begin{align}
U_{\mathrm{LEO}} &= \frac{GM_{\oplus}}{r_{\mathrm{LEO}}} + \frac{GM_{\mathrm{Moon}}}{|\mathbf{r}_{\mathrm{LEO}} - \mathbf{r}_{\mathrm{Moon}}|} + \frac{GM_{\odot}}{|\mathbf{r}_{\mathrm{LEO}} - \mathbf{r}_{\odot}|} \nonumber \\
&\quad + \frac{GM_{\oplus}}{r_{\mathrm{LEO}}}\sum_{l=2}^{\infty}\left(\frac{R_{\oplus}}{r_{\mathrm{LEO}}}\right)^l \sum_{m=-l}^{l} C_{lm}Y_{lm}(\theta_{\mathrm{LEO}}, \phi_{\mathrm{LEO}}), \label{eq:ULEO} \\
U_{\mathrm{DRO}} &= \frac{GM_{\oplus}}{r_{\mathrm{DRO}}} + \frac{GM_{\mathrm{Moon}}}{|\mathbf{r}_{\mathrm{DRO}} - \mathbf{r}_{\mathrm{Moon}}|} + \frac{GM_{\odot}}{|\mathbf{r}_{\mathrm{DRO}} - \mathbf{r}_{\odot}|} \nonumber \\
&\quad + \frac{GM_{\oplus}}{r_{\mathrm{DRO}}}\sum_{l=2}^{4}\left(\frac{R_{\oplus}}{r_{\mathrm{DRO}}}\right)^l \sum_{m=-l}^{l} C_{lm}Y_{lm}(\theta_{\mathrm{DRO}}, \phi_{\mathrm{DRO}}), \label{eq:UDRO}
\end{align}
For circular orbits, the orbital speed $v$ is related to the orbital radius by $v=\sqrt{GM/r}$.

And the frequency ratio of Eq.~\eqref{eq:freqratio_full} can be decomposed into the following physically meaningful components:
\begin{align}
\frac{f_{\mathrm{LEO}}}{f_{\mathrm{DRO}}} = 1 + \Delta_{\mathrm{Earth}} + \Delta_{\mathrm{Moon}} + \Delta_{\mathrm{Sun}} + \Delta_{\mathrm{FD}} + \Delta_{\mathrm{Doppler}} + \Delta_{\mathrm{Nonlinear}} + \mathcal{O}(c^{-5}), \label{eq:decomp_final}
\end{align}
where each term is defined below.

The Earth's gravitational potential contribution is:
\begin{equation}
\Delta_{\mathrm{Earth}} = \frac{1}{c^2}\left(U_{\oplus}^{\mathrm{DRO}} - U_{\oplus}^{\mathrm{LEO}}\right), \label{eq:DeltaEarth_def}
\end{equation}
with
\begin{align}
U_{\oplus}^{\mathrm{DRO}} &= \frac{GM_{\oplus}}{r_{\mathrm{DRO}}}\left[1 + \sum_{l=2}^{4}\left(\frac{R_{\oplus}}{r_{\mathrm{DRO}}}\right)^l \sum_{m=-l}^{l} C_{lm}Y_{lm}(\theta_{\mathrm{DRO}},\phi_{\mathrm{DRO}})\right], \label{eq:UoplusDRO_def} \\
U_{\oplus}^{\mathrm{LEO}} &= \frac{GM_{\oplus}}{r_{\mathrm{LEO}}}\left[1 + \sum_{l=2}^{\infty}\left(\frac{R_{\oplus}}{r_{\mathrm{LEO}}}\right)^l \sum_{m=-l}^{l} C_{lm}Y_{lm}(\theta_{\mathrm{LEO}},\phi_{\mathrm{LEO}})\right]. \label{eq:UoplusLEO_def}
\end{align}
The lunar and solar tidal potentials are:
\begin{align}
\Delta_{\mathrm{Moon}} &= \frac{GM_{\mathrm{Moon}}}{c^2}\left(\frac{1}{|\mathbf{r}_{\mathrm{DRO}} - \mathbf{r}_{\mathrm{Moon}}|} - \frac{1}{|\mathbf{r}_{\mathrm{LEO}} - \mathbf{r}_{\mathrm{Moon}}|}\right), \label{eq:DeltaMoon_def} \\
\Delta_{\mathrm{Sun}} &= \frac{GM_{\odot}}{c^2}\left(\frac{1}{|\mathbf{r}_{\mathrm{DRO}} - \mathbf{r}_{\odot}|} - \frac{1}{|\mathbf{r}_{\mathrm{LEO}} - \mathbf{r}_{\odot}|}\right). \label{eq:DeltaSun_def}
\end{align}
The frame-dragging term is:
\begin{equation}
\Delta_{\mathrm{FD}} = \frac{2G}{c^4}\left(J_{\mathrm{tot}}\frac{h_z^{\mathrm{DRO}}}{r_{\mathrm{DRO}}^3} - J_{\oplus}\frac{h_z^{\mathrm{LEO}}}{r_{\mathrm{LEO}}^3}\right), \label{eq:DeltaFD_def}
\end{equation}
where $h_z = (\mathbf{r} \times \mathbf{v})_z$. For a high-inclination LEO orbit with $i_{\mathrm{LEO}} = 70^\circ$, the LEO term contributes the dominant signal of order $10^{-20}$; the DRO term is approximately three orders of magnitude smaller and serves as a consistency check.

The Doppler shift is:
\begin{equation}
\Delta_{\mathrm{Doppler}} = \frac{1}{2c^2}\left[\left(1 + \frac{2GM_{\oplus}}{c^2r_{\mathrm{DRO}}}\right)v_{\mathrm{DRO}}^2 - \left(1 + \frac{2GM_{\oplus}}{c^2r_{\mathrm{LEO}}}\right)v_{\mathrm{LEO}}^2\right], \label{eq:DeltaDoppler_def}
\end{equation}
and the nonlinear term is:
\begin{equation}
\Delta_{\mathrm{Nonlinear}} = -\frac{2G^2M_{\oplus}^2}{c^4}\left(\frac{1}{r_{\mathrm{LEO}}^2} - \frac{1}{r_{\mathrm{DRO}}^2}\right). \label{eq:DeltaNonlinear_def}
\end{equation}

The typical magnitudes of these contributions for the nominal orbital parameters are summarised in Table~\ref{tab:magnitudes}.

\begin{table}[htbp]
\centering
\caption{Order-of-magnitude estimates of the frequency shift components for the LEO-DRO configuration. The estimates are based on $r_{\mathrm{LEO}} \approx 7.0\times10^6$ m, $v_{\mathrm{LEO}} \approx 7.55\times10^3$ m/s, $r_{\mathrm{DRO}} \approx 3.0\times10^8$ m, $v_{\mathrm{DRO}} \approx 1.3\times10^3$ m/s.}
\begin{tabular}{lcc}
\hline
\textbf{Term} & \textbf{Typical expression} & \textbf{Magnitude} \\
\hline
$\Delta_{\mathrm{Earth}}$ & $\frac{GM_{\oplus}}{c^2}\left(\frac{1}{r_{\mathrm{DRO}}} - \frac{1}{r_{\mathrm{LEO}}}\right)$ & $-6.2\times10^{-10}$ \\
$\Delta_{\mathrm{Doppler}}$ & $\frac{1}{2c^2}\left(v_{\mathrm{DRO}}^2 - v_{\mathrm{LEO}}^2\right)$ & $-3.1\times10^{-10}$ \\
$\Delta_{\mathrm{Moon}}$ & $\frac{GM_{\mathrm{Moon}}}{c^2}\left(\frac{1}{|\mathbf{r}_{\mathrm{DRO}} - \mathbf{r}_{\mathrm{Moon}}|} - \frac{1}{|\mathbf{r}_{\mathrm{LEO}} - \mathbf{r}_{\mathrm{Moon}}|}\right)$ & $1.8\times10^{-12}$ \\
$\Delta_{\mathrm{Sun}}$ & $\frac{GM_{\odot}}{c^2}\left(\frac{1}{|\mathbf{r}_{\mathrm{DRO}} - \mathbf{r}_{\odot}|} - \frac{1}{|\mathbf{r}_{\mathrm{LEO}} - \mathbf{r}_{\odot}|}\right)$ & $5.7\times10^{-14}$ \\
$\Delta_{\mathrm{Nonlinear}}$ & $-\frac{2G^2M_{\oplus}^2}{c^4}\left(\frac{1}{r_{\mathrm{LEO}}^2} - \frac{1}{r_{\mathrm{DRO}}^2}\right)$ & $-8.0\times10^{-19}$ \\
$\Delta_{\mathrm{FD}}$ & $\frac{2GJ_{\oplus}}{c^4}\frac{v_{\mathrm{LEO}}\cos i_{\mathrm{LEO}}}{r_{\mathrm{LEO}}^2}$ & $10^{-20}$ ($i_{\mathrm{LEO}}$\ \ dependent) \\
\hline
\end{tabular}
\label{tab:magnitudes}
\end{table}

The estimates in Table~\ref{tab:magnitudes} show that the dominant contributions are the Earth's gravitational potential ($\sim10^{-10}$) and the kinematic Doppler shift ($\sim10^{-10}$). These are ten orders of magnitude larger than the target signal but are either constant or slowly varying. The lunar tidal term ($\sim10^{-12}$) and solar tidal term ($\sim10^{-14}$) vary on timescales much longer than the LEO orbital period. The nonlinear term is significantly larger than the target signal but is correlated with the satellite's position and can be accurately modeled. The frame-dragging signal of order $10^{-20}$ arises from the gravitomagnetic vector potential. Recent multiplexed optical lattice clocks have already demonstrated a fractional frequency instability of \(9.7(4)\times10^{-18}/\sqrt{\tau}\), reaching a fractional uncertainty of \(8.9\times10^{-20}\) after 3.3 hours of averaging~\cite{Zheng2022}. With such performance, the clock quantum projection noise falls below the target signal and makes the detection possible.

Consequently, after the two-way cancellation of the first-order Doppler shift by the bidirectional pulsed laser link, the main contributors to the frequency comparison are the gravitational potentials of the Earth, Moon, and Sun, superimposed on the tiny frame-dragging signal. The higher-degree spherical harmonics contribute measurably at LEO ($\sim10^{-13}$) but become negligible at DRO distances. Thus, the extraction of the frame-dragging effect reduces to precisely modeling and subtracting the known gravitational potentials, as discussed in Sec.~\ref{sec:systematics}.

\section{SYSTEMATIC ERROR SOURCES AND SUPPRESSION STRATEGIES FOR FRAME-DRAGGING DETECTION}
\label{sec:systematics}
Detecting a frame-dragging signal at the $10^{-20}$ level requires not only accurate modeling of gravitational potentials but also stringent control of instrumental and environmental noise.
Following the error-analysis methodology developed for the DRO-based gravitational redshift mission, we categorize the uncertainty sources into five groups: clock-related noise, link and propagation effects, environmental perturbations, gravitational-field modeling uncertainties, and orbit-determination errors. For each uncertainty source, we estimate its magnitude, discuss the corresponding mitigation strategy, and assess the improvements required to reach the $10^{-20}$ level.

{Clock noise and systematic uncertainties.}
Quantum noise constitutes one of the fundamental statistical limitations of optical lattice clocks. To detect the predicted frame-dragging signal, the differential clock instability must approach the $10^{-20}$ level over the relevant averaging time.
Spin squeezing or quantum non-demolition measurements can further push the noise floor below the standard quantum limit. The combined measurement uncertainty introduced by the optical frequency comb and the frequency comparison processing unit is estimated to be $1.1\times10^{-18}$, based on the demonstrated performance of free-space optical frequency comb transfer at the $10^{-19}$ level and the projected in-orbit performance of the optical clock frequency comparison system. Next-generation zero-dead-time and multiplexed lattice clocks have already demonstrated fractional instabilities at the \(10^{-19}\) level at 20,000~s and uncertainties of \(8.9\times10^{-20}\) after 3.3~hours~\cite{Liu2025arXiv,Zheng2022}.

{Link and propagation noise}. The bidirectional pulsed laser link cancels the first-order Doppler shift, as demonstrated by the DOWR technique in DRO-A. After cancellation, the residual Doppler uncertainty is dominated by the \(\mathcal{O}(c^{-2})\) terms in the two-way frequency comparison model. For the LEO--DRO inter-satellite link, the residual Doppler uncertainty is estimated to be \(5\times10^{-18}\), following the same bidirectional cancellation analysis as in the DRO redshift mission~\cite{Qin2025}. This can be further reduced to \(5\times10^{-19}\) if the line-of-sight ranging accuracy is improved to the sub-meter level using the optical frequency comb ranging capability. This residual Doppler uncertainty is significantly smaller than the value that would be obtained from a naive scaling of the \(\mathcal{O}(c^{-1})\) Doppler term, demonstrating the effectiveness of the bidirectional cancellation scheme. Residual errors also arise from asymmetric delays in the uplink and downlink paths, including relativistic Shapiro time delay variations. The Shapiro delay contribution, after ephemeris correction using DE440, is estimated to be \(1\times10^{-20}\)~\cite{Qin2025}, which is entirely negligible. Unlike ground--satellite links, the inter-satellite optical link is free from tropospheric and ionospheric disturbances, and the atmospheric delay contribution is therefore absent for the LEO--DRO frequency comparison. With phase-stabilised fiber or free-space delay lines, the link noise can be suppressed to below \(10^{-20}\) for integration times of practical interest. In the future, sub-meter to decimeter inter-satellite ranging, combined with improved two-way phase stabilization and real-time delay calibration, can push the residual link noise further below the target signal level.

{Environmental perturbations}. The optical clock frequency is sensitive to temperature and magnetic fields. The blackbody radiation shift and the quadratic Zeeman shift are mitigated by active thermal control ($\Delta T < 1$ mK per day) and multi-layer $\mu$-metal shielding (residual field $<1$ nT). In situ monitoring and feedforward correction may further reduce the residual frequency uncertainty to below $1\times10^{-20}$.

{Gravitational field modeling uncertainties}. The Earth's gravitational potential contribution is calculated from a high-resolution geopotential model (e.g., EGM2008 or EGM2020). It should be emphasized that modern static gravity field models are already expanded to very high degree and order: EGM2008 contains spherical harmonic coefficients up to degree and order 2190, and newer combined models such as EGM2020 and other ultra-high-degree solutions extend well beyond degree 1000. With centimetre-level orbital accuracy for the LEO satellite and the current Earth gravity field model accuracy (geoid height error \(\sim 20\)~cm), the raw Earth potential modeling error is \(2.18\times10^{-17}\) at the ground level~\cite{rummel2011goce,mcgirr2023}. However, through local gravity field calibration, this uncertainty can be reduced by one order of magnitude to \(2.2\times10^{-18}\)~\cite{Qin2025}, which is the value adopted in our error budget. More importantly, because the static field is now known to degree \(>1000\), the high-degree harmonic contribution at LEO is already modeled at the \(10^{-13}\) level or better, and the residual static-field error after calibration is dominated not by truncation but by orbit determination and time-variable gravity. At DRO distances, the lunar gravity field model uncertainty contributes less than \(2\times10^{-21}\)~\cite{Qin2025}. The Earth's time-varying gravity field contributes \(1.8\times10^{-18}\) at the ground-to-DRO level~\cite{Qin2025}; the lunar time-varying gravity field is negligible at the \(10^{-19}\) level. The lunar and solar tidal potentials are computed from DE440 ephemerides; their uncertainty contribution is \(1\times10^{-20}\)~\cite{Qin2025}. Higher-degree Earth harmonics (\(J_3, J_4\), etc.) and non-linear terms such as \(G^2M^2/c^4\) are either negligible at DRO distances or can be modeled with sufficient accuracy. Future gravity missions such as NGGM and MAGIC are projected to improve the spatial and temporal resolution of time-variable gravity fields to \(\sim150\)~km and sub-weekly scales, 
respectively~\cite{rummel2011goce,mcgirr2023,Floberghagen2011,Daras2025}, which will further reduce the residual gravity-field modeling error. With ultra-high-degree static models and improved time-variable gravity calibration, the gravitational field modeling contribution can be pushed below the \(10^{-20}\) level.

{Orbit determination errors}. For the LEO satellite at \(\sim630\)~km altitude, centimetre-level orbital accuracy (achievable via GNSS and SLR) and millimetre-per-second velocity accuracy limit the combined gravitational redshift and time-dilation modeling error. Based on the error propagation analysis for the DRO mission~\cite{Qin2025}, the orbit determination error contribution to the LEO--DRO frequency comparison is estimated to be \(2.8\times10^{-18}\), dominated by the velocity error contribution to the second-order Doppler (time-dilation) term. Detailed covariance analysis confirms that the residual orbit-error contribution is \(2.8\times10^{-18}\) for the DRO--LEO configuration. To reduce this term, a geocentric distance accuracy of 1~dm and a velocity accuracy of 0.1~mm/s~\cite{Chen2023} are already being approached by state-of-the-art GNSS, SLR, and inter-satellite ranging systems. With further improvements in real-time kinematic positioning, inter-satellite laser ranging, and reduced dynamical model errors, the orbit-determination contribution can be suppressed to the \(10^{-19}\) level or below, compatible with the target \(10^{-20}\) signal after averaging. Finally, for a fixed LEO inclination, the frame-dragging (FD) frequency shift is a constant offset (\(\propto \cos i_{\mathrm{LEO}}\)), hard to separate from other constant systematics. We therefore propose a modulation strategy in which the LEO orbital inclination $i_{\mathrm{LEO}}$ is periodically varied through controlled orbital maneuvers. Since the FD signal scales with \(\cos i_{\mathrm{LEO}}\), this induces a predictable variation in the FD contribution while leaving most instrumental and environmental biases unchanged, allowing the FD effect to be isolated by cross-correlation. This strategy will be explored in future mission design studies.

Based on the above analyses, the frame-dragging signal is of order \(10^{-20}\), a level that is becoming increasingly accessible with next-generation optical clocks. These technological developments motivate further investigation of the LEO--DRO architecture, particularly through detailed sensitivity modeling and systematic-error analysis.
\section{Conclusion}
In this work, we have developed a theoretical framework and measurement concept for detecting the frame-dragging effect through high-precision optical-clock comparisons between satellites in LEO and DRO.
Working within the post-Newtonian approximation in the Geocentric Celestial Reference System, we have derived
the relativistic frequency ratio between a LEO satellite and a DRO satellite, 
accounting for contributions from Earth's monopole and higher-order gravitational multipoles, lunar and solar tidal potentials, relativistic Doppler effects, and the gravitomagnetic contribution associated with frame dragging.
We find that the frame-dragging contribution to the fractional frequency shift is of order $10^{-20}$ for the orbital configuration considered. Resolving such a weak signal therefore requires stringent control of systematic uncertainties and accurate modeling of the much larger gravitational and kinematic contributions.

To isolate this weak signal, we have developed a differential frequency-comparison scheme based on a bidirectional pulsed-laser link between the LEO and DRO satellites.
An appropriate combination of the two-way measurements suppresses the leading first-order Doppler shift, while the residual kinematic contributions can be modeled and subtracted.
The dominant gravitational contributions from the Earth, Moon, and Sun are evaluated using a high-resolution terrestrial geopotential model and the JPL DE440 planetary and lunar ephemerides. 
Under the assumed clock specifications and integration time, the quantum projection noise is expected to fall below the $10^{-20}$ level. Provided that the residual systematic uncertainties and link noise are sufficiently controlled, this sensitivity would enable the detection of the predicted frame-dragging signal.

Achieving the sensitivity required for this measurement will rely on several demanding technological capabilities, including next-generation optical lattice clocks with a projected fractional frequency instability of $\sigma_y(\tau)=10^{-20}/\sqrt{\tau}$, thermally stabilized optical reference cavities ($\Delta T<1~\mathrm{mK}$), magnetic shielding ($<1\,\mathrm{nT}$), and a drag-free platform (residual acceleration $<1\times10^{-9}\,\mathrm{m/s^2}$). 
And further improvements in clock stability would not only strengthen the frame-dragging measurement but also open new possibilities for mapping Earth's mass transport, such as ocean currents and mantle convection, via the relativistic gravitomagnetic vector potential. Moreover, the proposed LEO-DRO architecture provides a natural path for future tests of general relativity in the cislunar regime, 
including precision gravitational-redshift measurements and searches for violations of local position invariance.

Although our results indicate that next-generation optical clocks in the LEO--DRO configuration could achieve the sensitivity required to detect the frame-dragging effect, the leading contribution for a nearly circular LEO orbit with a fixed inclination $i_{\mathrm{LEO}}$ appears as an approximately constant frequency offset ($\Delta_{\mathrm{FD}}\propto\cos i_{\mathrm{LEO}}$). This may complicate its separation from stationary instrumental biases. A modulation strategy could help distinguish the frame-dragging contribution through cross-correlation of the measured frequency residuals with the predicted signal template. Inclination-dependent gravitational and kinematic backgrounds would also need to be accounted for. The feasibility of this approach will be further explored in future mission studies.

Frame dragging affects not only orbital but also the proper-time evolution of moving clocks through the gravitomagnetic metric components $g_{0i}$~\cite{Mashhoon:2001ka,Hackmann:2014aga}. These components also contribute to the path dependence of Einstein clock synchronization, which can lead to a nonzero synchronization gap around a closed spatial loop~\cite{Bini:2000ge, Costa:2021atq}. The proposed LEO--DRO measurement therefore provides an independent test of terrestrial gravitomagnetism, with implications for future experiments probing the nontransitivity of clock synchronization.

In summary, by combining a differential frequency-comparison scheme with a bidirectional optical link, our approach provides an independent observational strategy for measuring terrestrial frame dragging.
By combining a differential frequency-comparison scheme with a bidirectional laser link, our approach provides an independent observational strategy for measuring terrestrial frame dragging. The relativistic frequency-transfer analysis, sensitivity estimates, and instrumental requirements presented in this work identify the conditions necessary for a future clock-based detection of the frame-dragging effect. These results establish a foundation for the development of next-generation spaceborne experiments at the intersection of quantum metrology and fundamental gravitational physics.

\begin{acknowledgments}
This work is supported by the National Key R\&D
Program of China under Grants No. 2025YFE0217300, the National Natural Science Foundation of China (12575069 and 12475049), the International Partnership Program of the Chinese Academy of Sciences Grant No. 025GJHZ2023106GC, the China National Key Research and Development Program No. 2024YFC2207300, the Strategic Priority Research Program on Space Science,
the Chinese Academy of Sciences (XDA30040000, XDA30030000). This work is also supported by the High-performance Computing Platform of Peking University.
\end{acknowledgments}

\appendix
\section{ORDER-OF-MAGNITUDE ANALYSIS AND TRUNCATION OF HIGHER-ORDER TERMS}
\label{app:orders}
The full expression in Eq.~\eqref{eq:freqratio_full} contains the relevant effects up to $\mathcal{O}(c^{-4})$. However, to keep the subsequent analysis tractable while preserving physical rigor, we must identify which terms are numerically significant for a measurement at the $10^{-20}$ level. We examine the individual contributions below.

{Earth's monopole and Doppler terms.} The Earth's gravitational potential difference $\Delta_{\mathrm{Earth}} = (GM_{\oplus}/c^2)(1/r_{\mathrm{DRO}} - 1/r_{\mathrm{LEO}}) \approx 6.2\times10^{-10}$ and the Doppler shift $\Delta_{\mathrm{Doppler}} \approx (v_{\mathrm{LEO}}^2 - v_{\mathrm{DRO}}^2)/(2c^2) \approx 3.1\times10^{-10}$ are the dominant contributions. Both are static or vary slowly on timescales much longer than the LEO orbital period. These terms are accurately modeled to avoid systematic biases.

{Earth's higher-order spherical harmonics.} The leading \(J_2\) term at LEO contributes \(\sim5.7\times10^{-13}\), with higher-degree harmonics \((J_3,J_4,\ldots)\) contributing cumulatively at the \(\sim10^{-13}\) level. At DRO distances, these harmonics are suppressed by successive powers of \(R_\oplus/r_{\rm DRO}\approx2\times10^{-2}\) and become negligible for the required precision. We retain harmonics up to a sufficiently high degree in \(U_{\mathrm{LEO}}\) and truncate at \(l=4\) in \(U_{\mathrm{DRO}}\).

{Lunar tidal potential.} The lunar tidal contribution is \(\Delta_{\mathrm{Moon}}\sim1.8\times10^{-12}\) at DRO periapsis, varying on a \(\sim14\)-day synodic period. It is modeled and subtracted using DE440 ephemerides over the mission lifetime.

{Solar tidal potential.} The solar tidal contribution is $\Delta_{\mathrm{Sun}} \sim 5.7\times10^{-14}$, varying on diurnal and annual timescales. It is a small but non-negligible systematic effect for $10^{-18}$-level precision and is therefore retained, computed from DE440 ephemerides.

{Gravitomagnetic (frame-dragging) term.} This is the target signal. The Earth's gravitomagnetic contribution is:
\begin{equation}
\Delta_{\mathrm{FD}}^{(L)} = \frac{2G J_{\oplus}}{c^4}\frac{v_{\mathrm{LEO}} \cos i_{\mathrm{LEO}}}{r_{\mathrm{LEO}}^2} \sim 10^{-20}\cos i_{\mathrm{LEO}},
\end{equation}
which is precisely the signal we aim to detect. The DRO contribution is approximately three orders of magnitude smaller and is included for completeness. 

{Nonlinear terms ($-2W^2/c^4$).} The leading nonlinear contribution, arising from the Earth's self-gravity, is:
\begin{equation}
\Delta_{\mathrm{Nonlinear}} = -\frac{2G^2M_{\oplus}^2}{c^4}\left(\frac{1}{r_{\mathrm{LEO}}^2} - \frac{1}{r_{\mathrm{DRO}}^2}\right) \approx -8.0\times10^{-19}.
\end{equation}
It can be modeled accurately and is therefore retained. 

{Sun's gravitomagnetic contribution.} The Sun's orbital motion produces a gravitomagnetic field that, at Earth's location, has a fractional frequency effect of order:
\begin{equation}
\Delta_{\mathrm{Sun,GM}} \sim \frac{G J_{\odot}}{c^4 r_{\odot}^2} v_{\oplus} \sim 10^{-27},
\end{equation}
which is seven orders of magnitude below the target signal. This term is safely neglected. Similarly, the Moon's gravitomagnetic contribution at LEO is $\sim10^{-24}$ and is neglected, while at DRO it contributes at the $\sim10^{-24}$ level and is included through $J_{\mathrm{tot}}$ in Eq.~\eqref{eq:freqratio_full}.

Spatial-metric contributions. The terms arising from $g_{ij}$ in the Taylor expansion produce corrections of relative order $(v/c)^2$ compared to the corresponding $g_{00}$ terms. For the Earth monopole, this gives an additional contribution:
\begin{equation}
\frac{1}{2c^2}\left[\frac{2GM_{\oplus}}{c^2r_{\mathrm{LEO}}}v_{\mathrm{LEO}}^2 - \frac{2GM_{\oplus}}{c^2r_{\mathrm{DRO}}}v_{\mathrm{DRO}}^2\right] \sim 10^{-19},
\end{equation}
which is comparable to the target signal. This contribution is already included in the Doppler term in Eq.~\eqref{eq:freqratio_full} through the $(1 + 2GM_{\oplus}/(c^2r))$ factors. The contributions from the lunar and solar tidal potentials to $g_{ij}$ are suppressed by a factor of order $(v/c)^2$ relative to the corresponding tidal-potential terms. Their estimated magnitudes are $\lesssim10^{-22}$ and are therefore negligible.

Based on this magnitude analysis, the only terms that are retained in the final model are those that contribute at levels $\gtrsim 10^{-20}$ in fractional frequency. Table~\ref{tab:magnitudes} summarizes the retained contributions and their estimated magnitudes. Terms that are below the $10^{-20}$ threshold---specifically, the Sun's gravitomagnetic contribution ($\sim10^{-27}$), the Moon's gravitomagnetic contribution at LEO ($\sim10^{-24}$), the Earth--Moon and Earth--Sun cross nonlinear terms ($\sim10^{-22}$), and the tidal contributions to $g_{ij}$ ($\lesssim10^{-22}$)---are omitted from the subsequent analysis. This truncation is justified by their negligible impact on a measurement with target sensitivity $10^{-20}$.

\bibliography{ref}

\end{document}